\documentclass[sigconf,natbib=true,anonymous=false,nonacm]{acmart}

\usepackage{xspace}
\usepackage{multirow}
\usepackage{graphics}
\usepackage{amsfonts}
\usepackage{subcaption}
\usepackage{graphics}
\usepackage{enumitem}
\usepackage{float}

\begin{document}

\newcommand{\ourdataset}{\texttt{SE-PQA}\xspace} 
\newcommand{\ssp}{\hphantom{*}} %for StarSPace in tables

\title{\ourdataset: Personalized Community Question Answering}

\author{Pranav Kasela}
\orcid{0000-0003-0972-2424}
\affiliation{
  \institution{University of Milano-Bicocca}
  \city{Milan}
  \country{Italy}
}
\affiliation{
  \institution{ISTI-CNR, Pisa, Italy}
   \city{Pisa}
   \country{Italy}
}
\email{p.kasela@campus.unimib.it}

\author{Marco Braga}
\orcid{0009-0004-7619-8399}
\affiliation{
  \institution{University of Milano-Bicocca}
  \city{Milan}
  \country{Italy}
}
\affiliation{
  \institution{Politecnico di Torino, DAUIN}
  \city{Turin}
  \country{Italy}
}
\email{m.braga@campus.unimib.it}

\author{Gabriella Pasi}
\orcid{0000-0002-6080-8170}
\affiliation{
  \institution{University of Milano-Bicocca}
  \city{Milan}
  \country{Italy}
}
\email{gabriella.pasi@unimib.it}

\author{Raffaele Perego}
\orcid{0000-0001-7189-4724}
\affiliation{
  \institution{ISTI-CNR, Pisa, Italy}
   \city{Pisa}
   \country{Italy}
   }
 \email{raffaele.perego@isti.cnr.it}

\renewcommand{\shortauthors}{Kasela, et al.}

\begin{abstract}
Personalization in Information Retrieval is a topic studied for a long time.
Nevertheless, there is still a lack of high-quality, real-world  datasets to conduct large-scale experiments and evaluate models for personalized search.
This paper contributes to filling this gap by introducing \ourdataset (StackExchange - Personalized Question Answering), a new curated resource to design and evaluate personalized models related to the task of community Question Answering (cQA). The contributed dataset  includes  more than  1 million queries and 2 million answers,  annotated with a rich set of features modeling the social interactions  among the users of a popular cQA platform. We describe the characteristics of \ourdataset and detail the features associated with questions and answers. We also provide reproducible baseline methods for the cQA task based on the resource, including  deep learning models and personalization approaches. The results of the preliminary experiments conducted 
show the appropriateness of \ourdataset to train effective cQA models; they also show that personalization remarkably improves the effectiveness of all the methods tested. Furthermore, we show the benefits in terms of robustness and generalization of combining data from multiple communities for personalization purposes. 

\end{abstract}

\maketitle

\section{Introduction}

Personalization is a problem studied for a long time in Information Retrieval (IR) ~\cite{perIR1,perIR2,perIR3,perIR4,perIR5, perIR6} and Natural Language Processing (NLP) ~\cite{perIR6}.
Personalized search aims to tailor the search outcome to a specific user (or group of users) based on the knowledge of her/his interests and online behaviour.
Given the ability of Deep Neural Network (DNN) models to face many different tasks by extracting relevant features from both texts and structured sources~\cite{bhaskar_nir}, there is the expectation of a huge potential also for their application in Personalized IR (PIR) and Recommender Systems (RS).
However, the lack of publicly-available, large-scale datasets that include user-related information is one of the biggest obstacles to the training and evaluation of DNN-based personalized models.
Some real-world datasets are commonly used in the literature to design and assess personalization models. These datasets include the AOL query log~\cite{aol}, the Yandex query log\footnote{https://www.kaggle.com/c/yandex-personalized-web-search-challenge}, and the CIKM Cup 2016 dataset\footnote{https://competitions.codalab.org/competitions/11161}. Moreover, even  synthetically enriched datasets have been used such as: PERSON~\cite{person}, the Amazon product search dataset~\cite{amazon_search_dataset}, and a dataset based on the Microsoft Academic Knowledge Graph~\cite{mag_dataset}.
However, all of them have some issues. For example, ethical and privacy issues are related to using the AOL query log~\cite{aol_privacy}. In contrast, the anonymization performed on the Yandex query log prevents its use for training or fine-tuning natural language models.

This paper aims to fill this gap by contributing \ourdataset (StackExchange - Personalized Question Answering), a large dataset rich in user-level features that can be exploited  for training and evaluating personalized models addressing the \textit{community Question Answering}  (cQA) task. 
\ourdataset is based on StackExchange\footnote{\url{https://stackexchange.com}}, a  popular cQA platform with a network of $178$ open forums. A dump of the StackExchange user-contributed content is publicly available\footnote{\url{https://archive.org/details/stackexchange}} according to a cc-by-sa 4.0 license\footnote{\url{https://archive.org/details/stackexchange}} according to a cc-by-sa 4.0 license\footnote{There are although some disputes concerning a unilateral change of licensing policy  by StackExchange: \url{https://meta.stackexchange.com/questions/333089/stack-exchange-and-stack-overflow-have-moved-to-cc-by-sa-4-0}}.
With great care, we have preprocessed the original dump by building \ourdataset, a curated dataset with
about one million questions and two million associated answers annotated with a rich set of features modeling the social interactions of the user community.  The features include, for example,  the positive or negative votes received by a question or an answer, the number of views, the number of users that selected a given question as a favorite one, the tags from a controlled folksonomy describing the topic dealt with, the comments that other users might have written under a question or an answer. To favor the design and evaluation of personalized models,  the users in \ourdataset are associated with their past questions and answers, their social autobiography, their reputation score, and the number of views  received by their profile.
The cQA task can be addressed on \ourdataset with different methodologies exploiting either the textual description of questions and answers, the folksonomy, the features modeling the social interactions, or a combination of the above information sources. In this paper, we  focus on IR approaches to cQA. Thus, we adapt the cQA task  to an ad-hoc retrieval task where the question is seen as a query, and the answers are retrieved from the pool of past answers indexed for the purpose.
In this particular setting, the system aims to retrieve a (small) ranked set of documents that contain the correct answers to the user question. 
There can be multiple correct answers given a question, so, in this case, personalization can be used to understand the user's context and background and rank higher the answers that are more relevant to the specific user.\newline
In summary, the novel contribution of this paper is the following:
\begin{itemize}
    \item We contribute  the \ourdataset dataset, a novel public resource consisting of a comprehensive corpus including 
    more than one million questions and two million answers by about $600k$ users. The richness and variety of features provided with the dataset enable its use for the design and evaluation of both classical and personalized cQA.
    
    \item We provide a detailed analysis of the resource made available in \ourdataset compared to those previously available and used in the research community.
    
    \item We report a preliminary comparison of the performance of different methods for cQA applied to the questions, answers, and users in \ourdataset. The results show that models based on deep learning outperform in  effectiveness traditional retrieval models, and that by exploiting personalization features we can obtain a significant performance boost.
\end{itemize}
The paper is organized as follows. 
Section \ref{sec:dataset} introduces the \ourdataset dataset and reports some statistics about its content. Furthermore, the section details the personalized cQA tasks addressed in this paper by using   \ourdataset. Moreover, it provides a comparison of \ourdataset with respect to other publicly-available resources in the  field.  
Section \ref{sec:experiments} presents a preliminary comparison of traditional and personalized models for cQA applied to \ourdataset. 
In Section \ref{sec:utility} we discuss the utility and the practical implications of the new resource.  
Finally, Section \ref{sec:conclusion} concludes the work and draws some future lines of investigation.

\section{The \ourdataset Dataset}
\label{sec:dataset}

The textual posts in StackExchange forums are associated with rich social metadata information. When users ask a question to the community, they assign some tags specifying the topic to make the question searchable and visible to the users interested in it. 
The questions are up-voted or down-voted by the community based on their interest and  adherence to the community 
guidelines\footnote{\url{https://meta.stackexchange.com/help/how-to-answer}}. In many cases, the community suggests to the question author how to improve the question if it is poorly expressed or formatted.
Similar treatment is given to the answers, which can be up-voted or down-voted by the community; moreover, the user who asked the question can also choose the answer he/she deems the best, which may differ from the one that received the most up-votes from others.  We note, however, that 87.6\% questions and  answers are  assigned a score given by the difference between the number of up and down-votes. 
A positive score thus indicates that the post has more up-votes than down-votes, while a negative score indicates that more users down-voted it.

\begin{figure}
    \centering
    \includegraphics[width=\linewidth]{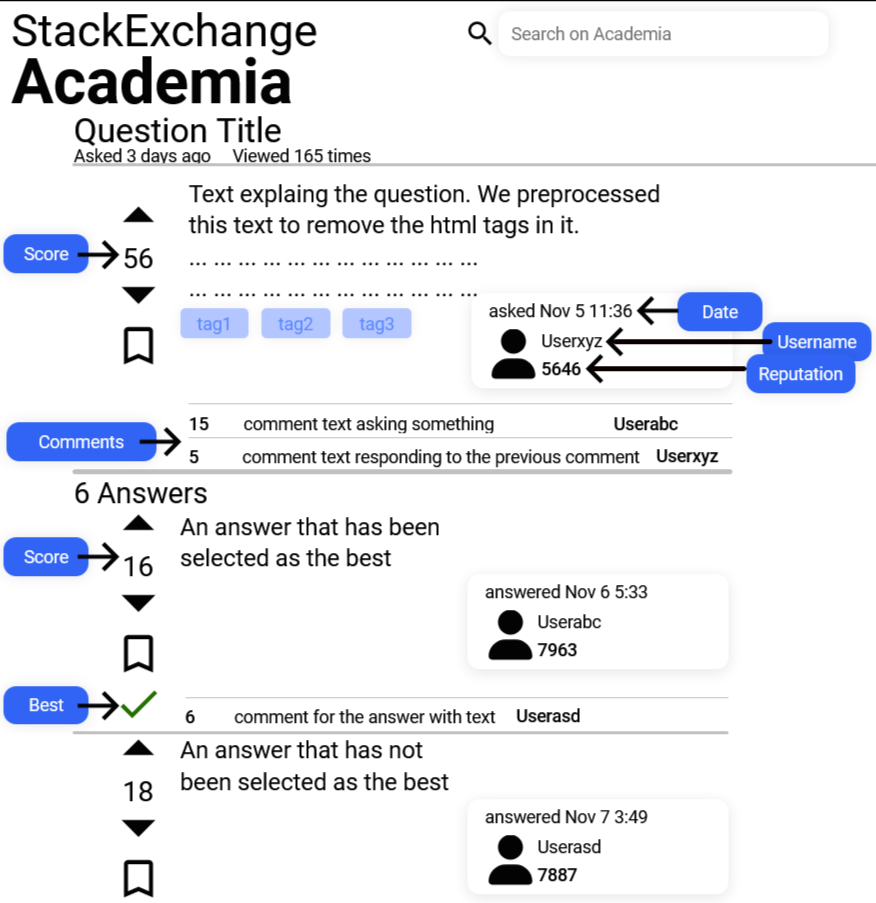}
    \caption{Illustration of StackExchange data.}
    \Description{UI example of the StackExchange website. The figure highlights the position of some of the most important features available, for example, post score, comment location, date of publication, username.}
    \label{fig:stack_ui}
\end{figure}

StackExchange  is quite  well known in the IR community: for example, it has been used for training  a language model for sentence similarity~\cite{huggingface_all}.
To the best of our knowledge, the usage of StackExchange for Q\&A tasks has been, however, limited just to selecting similar sentence training pairs without exploiting  user-level/social  features for  personalized information retrieval tasks. 
Another study uses  StackExchange  for  duplicate question retrieval~\cite{cqadupstack}. The dataset built for this task consists of 12 separate  communities, and the authors do not address the de-duplication task across  community boundaries.

With \ourdataset we overcome the previous limitations and provide a complete, curated dataset of textual questions and answers belonging to different, heterogeneous forums. 
In \ourdataset a user can belong to multiple communities; if we take into consideration users that wrote at least 2 questions, about 50\% of them have asked questions in multiple communities. As we increase the minimum number of questions, also the percentage of users using multiple communities increases. For instance, if we take only the users that wrote at least 5 documents (either questions or answers) and consider both the questions and the answers written by the user, we note that of the resulting 62k users only 23k (37\%) wrote either a question or an answer in only a single community, while 40k (63\%) wrote documents in at least two different communities, 26k (42\%) in at least three and 18k (28\%) in more than three communities.

We claim that personalization is particularly useful for multi-domain collections, where we can exploit information about users' interests in multiple topics of different domains; when the data is instead derived from a single domain or a specific topic of a domain, personalization may  become less important.  
We provide evidence  of this assertion in Section \ref{subsec:exResults} where we report about our experiments applying the same personalization approach to both the  complete \ourdataset dataset and to the data sampled from separate communities: the results show that personalization on the multi-domain dataset yields better improvements than on only single communities.\\
To increase diversity, in \ourdataset we thus combine data from multiple networks that can be categorized  under the large  umbrella of humanistic communities. These communities focus on different to
pics, but the language used is not too diverse among them. In particular, we choose the following 50 communities: \\

\noindent
\textit{writers, workplace, woodworking, vegetarianism, travel, sustainability, sports, sound, skeptics, scifi, rpg, politics, philosophy, pets, parenting, outdoors, opensource, musicfans, music, movies, money, martialarts, literature, linguistics, lifehacks, law, judaism, islam, interpersonal, hsm, history, hinduism, hermeneutics, health, genealogy, gardening, gaming, freelancing, fitness, expatriates, english, diy, cooking, christianity, buddhism, boardgames, bicycles, apple, anime, academia}.
\\

The training, validation, and test split are done temporally to avoid any kind of data leakage. The training set includes all questions written from \textit{2008-09-10} to \textit{2019-12-31} (included), the validation set is formed by questions asked between \textit{2019-12-31} and \textit{2020-12-31} (included), while the test set contains the questions from \textit{2019-12-31} till \textit{2022-09-25} (included).

\begin{figure}[t]
    \centering
    \includegraphics[width=\linewidth]{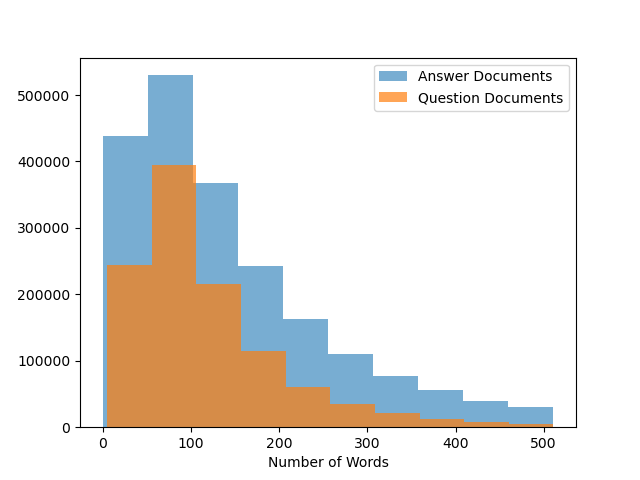}
    \caption{Word length distribution for questions and answers in \ourdataset.}
    \Description{The plot shows the distribution of questions and answer posts, in particular, how they are skewed to the left, which indicates that there are many posts with few words while in some cases the posts can be longer.}
    \label{fig:doc_word_hists}
\end{figure}

There are a total of $1,125,407$ questions in the dataset, $1,001,706$ of which have at least one answer ($89\%$ of all questions) and $525,030$ of which have a response that the questioner has selected as the best one ($47\%$ of all questions). We are left with 822 974 training questions, 78 854 validation questions, and 99 878 test questions after the temporal splits.\\
There are $2,173,139$ answers and $588,688$ users.
Many users in the communities register themselves just for asking a question and then never use their accounts again. In fact, the dataset has a median of 1 user-generated document (either a question or an answer), with about 80\% of users having no more than 2 documents.
The text  in the dataset is  preprocessed by removing HTML tags present in the original documents.
In Table \ref{tab:words_in_documents} we report the basic statistics for the dataset. Specifically: \textit{document length}, measured in the number of words, \textit{document score}, which is the difference between the number of up- and down-votes assigned by the community; \textit{answers' count}, the number of answers given to a question; \textit{comments' count},  the number of user comments to a given question or answer; \textit{favorite count}, that indicates the number of users that flagged the question as their favorite, showing their interest in that topic; \textit{tags count},  the number of tags associated to the question by the asking user.
From Table \ref{tab:words_in_documents} and Figure \ref{fig:doc_word_hists} we can notice that, as expected, most documents are short, with answers generally longer than questions.
The dataset is available at Zenodo\footnote{\url{https://doi.org/10.5281/zenodo.10679181}}.
The code to reproduce the dataset and the baseline is publicly available\footnote{\url{https://github.com/pkasela/SE-PQA}}.

\begin{table}[t]
    \centering
    \caption{Basic feature statistics for \textbf{ }question and answers.}
    \resizebox{\linewidth}{!}{
    \begin{tabular}{c|l|c|c|c|c|c|c}
        \toprule
        & Type & mean & std & median & 25\% & 75\% & 99\% \\
        \midrule
        \multirow{2}{*}{Questions} & Length & 125.69 & 112.90 & 94 & 60 & 153 & 553 \\
        & Score & 5.13 & 10.73 & 2 & 1 & 6 & 45 \\
        \#docs = & Answers Count & 1.93 & 1.92 & 1 & 1 & 2 & 9 \\
        1,125,407 & Comments Count & 2.78 & 3.37 & 2 & 0 & 4 & 15 \\
        & Favorite Count & 2.07 &	4.88 & 1 & 1 & 2 & 16 \\
        & Tags Count & 2.45 & 1.21 & 2 & 1 & 3 & 5 \\
        \midrule
        Answers & Length & 178.15 & 210.09 & 117 & 61 & 218 & 1000 \\
        \#docs = & Score & 5.13 & 12.43 & 2 & 1 & 5 & 51 \\ 
        2,173,139 &  Comments Count & 1.62 & 2.62 & 1 & 0 & 2 & 12\\
    \end{tabular}
    }
    \label{tab:words_in_documents}
\end{table}

\subsection{Task Definition}
\label{subsec:taskdefinition}

Even though \ourdataset can be used for many IR tasks (e.g., duplicate and related question retrieval or expert finding), we address here the cQA task only, by illustrating how it can be addressed by using the resources in \ourdataset.
The addressed cQA task focuses on satisfying the information needs expressed in user questions by retrieving relevant documents from a collection of historical answers posted by the  community members. 
We infer the relevance of an answer to a question from the number of up-votes given by community members.  Concerning the experiments involving personalized  cQA models, we only consider relevant the single answer that is explicitly labeled as the best answer by the user who submitted the question. 
More formally we provide the following definition.
Let $\mathcal{A}$ be a set of answers $\{a_1, \ldots, a_n\}$ posted by the members of the community and let $\mathbf{q}$ be a question asked by the user $\mathbf{u}$.
The objective of cQA is to retrieve a ranked list of $k$ answers $\{a_{q,1},\ldots,a_{q,k}\}$ from  $\mathcal{A}$ based on their relevance to $\mathbf{q}$. 
In our experiments with \ourdataset we infer the relevance of an answer to a question from the up-votes given by community members. With regards to the experiments involving personalized  cQA models, we consider instead only the answer $a_{q,u} \in \mathcal{A}$ that is the most relevant for the question $\mathbf{q}$ and for the specific user $\mathbf{u}$. This can be assessed on \ourdataset by considering the single answer that is explicitly labelled as the best answer by the user who submitted the question.

In order to address the above-defined cQA task, we preprocess the collection of  answers of \ourdataset. 
Specifically, we discard answers with negative scores since they are assumed to be of low quality and not relevant to the cQA task. 
This cleaning step affects about  100k answers. As a result, $2,073,370$ answers are left in the dataset.
Moreover, we discard all the questions  that have not received an answer.

To create the set of relevant answers  for the questions, \textit{i.e., the golden standard}, we consider the answers given to each question. \newline
A total of 525,030 questions out of 1,001,706 have an answer selected as the best by the user who asked the question. 
We are sure that this answer received  a positive score from the community since we have removed all the answers with negative scores. 
Thus, the answer given to a question $\mathbf{q}$ of user $\mathbf{u}$ and selected as best can be considered as both relevant to the question $\mathbf{q}$ (positive score from the community) and to the user $\mathbf{u}$ (selection of the best answer).\\
By using this information, we define two versions of this dataset: the \textit{base} version,  where we consider as relevant for a question all the answers having a positive score, and the personalized (\textit{pers}) version, which, instead, considers relevant for both the user and the question only the single answer that the user selected as the best answer.
We note that both versions of the dataset, \textit{base} and \textit{pers}, can be used for personalized cQA, with the following difference: in the \textit{pers} version each query is potentially personalizable, while the \textit{base} version also includes queries that cannot be always used to train personalized models since the choices of the answers preferred by the users are not always available.

A variety of user-generated information from the training set can be used in the personalization phase. 
For each question, we include all the user posts (questions and answers of the user asking the question) that were written prior to the question being asked.
This is done to avoid any data leakage for query-wise training, but the user data is not limited to these documents; in fact, one can also consider the social interaction between users, the tags assigned by the users to the previous questions asked along with their meaning, the badges earned by users. Furthermore, the dataset includes the biographic text (\textit{about me}) self-introducing each user,  a rich set of numeric features (e.g., user reputation score, number of up-votes and down-votes of each post,  number of views), a set of temporal information (e.g., user creation date, last access date, post creation timestamp). 

\subsection{Comparison with available datasets}

\begin{table*}[t]
\caption{Comparison between \ourdataset and other text-based datasets for personalized IR. All the datasets  are in English. %%.
}
\centering
\resizebox{\textwidth}{!}{
\begin{tabular}{l|c|c|c|c|c|c|c|c}
    \toprule
     Dataset & Documents & Train Queries & Val Queries & Test Queries & Users & Avg. user docs & Avg. \# relevant & Rel. Assessment\\
     \midrule
     \midrule
     AOLIA\cite{aolia} & 1 291 695 & 212 386 & 31 064 & 36 052 & 30 166 & 136.62 $\pm$ 134.17 & 1.15 $\pm$ 0.46 & inferred from click\\
     \midrule
     PERSON\cite{person} & 616 889 & - & - & - & 558 898 & - & 5.5 $\pm$ 5.3 & inferred from citation \\
     \midrule
     MAG Computer Science\cite{mag_dataset} & 4 809 684 & 552 798 & 5 583 & 6 497 & 5 260 279 & 61.94 $\pm$ 60.32 & 3.25 $\pm$ 3.27 & inferred from citations\\
     MAG Physics & 4 926 753 & 728 171 & 7 355 & 6 366 & 5 835 016 & 60.98 $\pm$ 56.54 & 4.17 $\pm$ 4.15 & inferred from citations\\
     MAG Political Science & 4 814 084 & 162 597 & 1 642 & 5 715 & 6 347 092 & 40.64 $\pm$ 29.32 & 3.88 $\pm$ 5.17 & inferred from citations\\
     MAG Psychology & 4 215 384 & 544 882 & 5 503 & 12 625 & 4 825 578 & 61.66 $\pm$ 62.72 & 4.73 $\pm$ 4.4 & inferred from citations\\
     \midrule
     Amazon Electorincs\cite{amazon_search_dataset} & 1 689 188 & 904 & - & 85 & 192 403 & 8.78 $\pm$ 8.26 &1.12 $\pm$ 0.48 & synthetic\\
     Amazon Kindle Store & 982 618 & 3 313 & - & 1 290 & 68 223 & 35.65 $\pm$ 37.48 & 1.87 $\pm$ 3.3 & synthetic\\
     Amazon CDs & 1 097 591 & 534 & - & 160 & 75 258 & 21.75 $\pm$ 16.53 & 2.57 $\pm$ 6.59 & synthetic\\
     Amazon Cell Phones & 194 439 & 134 & - & 31 & 27 879 & 4.95 $\pm$ 2.6 & 1.52 $\pm$ 1.13 & synthetic\\
     \midrule
     \ourdataset [Base] & 2 073 370 & 822 974 & 78 854 & 99 878 & 588 688 & 34.71 $\pm$ 107.33  & 2.07 $\pm$ 1.81 & inferred from answer scores\\
     \ourdataset [Pers] & 2 073 370 & 224 366 & 18 086 & 19 811 & 17 963 & 75.67 $\pm$ 142.41 & 1 & manually selected by user\\
     \bottomrule
\end{tabular}
}
\label{tab:pir_compare}
\end{table*}

We survey other works contributing datasets for personalized IR and discuss their limitations. These datasets cover a huge variety of tasks ranging  from web search to product search and academic search.

In Table \ref{tab:pir_compare} we summarize the basic statistics of the main  datasets used in the literature for personalized IR tasks.\\
The AOL query log was released in 2006 and even after the harsh criticism it received due to privacy-related issues, it remains to this date a widely employed resource for a variety of tasks, especially personalized ad-hoc retrieval. This dataset includes about 20 million Web queries issued by more than 657,000 users over three months (from 03/01/2006 to 05/31/2006).  For each query, the dataset details provide the userid, the URL, and the rank of the web page clicked, if any.
A huge limitation of this dataset is that the web pages in the corpus are represented by their URLs and the text is not provided. 
To cope with this issue, researchers use a version of the corpus collected in 2017~\cite{aol17} scraping the text content of the web pages in the corpus. This additional dataset  comes however with another problem: the content of web pages can change over time, and many documents might have changed from 2006 to 2017, thus  making the dataset less reliable. Recently, a new version of the dataset was proposed, which used Internet Archive to retrieve documents as they were in 2006~\cite{aolia}.
The AOLIA dataset~\cite{aolia} is a derivative of the original dataset (AOL Query Log~\cite{aol}); it has been cleaned to generate a higher-quality query set. First, queries with no clicks (and consequently no relevant documents) are removed. 
Then, all the queries with domain references (.com, .org, etc.) and queries pointing to adult or illegal websites are eliminated. Furthermore, all queries with fewer than three characters are discarded as well.
Finally, queries from users with less than twenty associated queries are removed in order to have enough user-related data to perform personalization. As a result, the AOLIA dataset contains about 1,3M documents and around 30k different users.\newline
To tackle the issues that come with the real words datasets, some synthetic datasets, and associated evaluation frameworks have been proposed in recent years: PERSON~\cite{person}, Amazon product search~\cite{amazon_search_dataset} and MAG, a dataset based on the Microsoft Academic Knowledge Graph~\cite{mag_dataset}.\newline
Tabrizi et al.~\cite{person} proposed PERSON as a synthetic personalized evaluation framework for IR based on citation networks. The authors base the dataset on the ArnetMiner citation network~\cite{aminer}. The idea is that, from the authors' perspective,  the  papers referenced in a document are somehow related to the document it is cited in. From an IR point of view, the document content (title or abstract) are considered as the user query, while the cited documents are assumed relevant to the query.\\ 
The dataset based on the Microsoft Academic Graph (MAG)~\cite{mag_dataset} follows a procedure similar to the one used by Tabrizi et. al; it uses a much larger citation graph to derive four different datasets, one for each of the following subjects: Political Science, Psychology, Computer Science and Physics.
Similarly to the PERSON framework,  queries come from paper titles, and the previous papers of a user are used to build her personal profile.
Paper titles from users authoring  less than twenty previous papers are removed from the dataset, providing the user-related information necessary  to define appropriate user models.
As explained in \citet{person},  such a dataset can be employed to develop and compare various user models, but cannot be used to assess  personalized search effectiveness due to the strong assumptions made to determine relevant documents in this framework.

The Amazon product search dataset~\cite{amazon_search_dataset} is based on the Amazon Review dataset~\cite{amazon_core_dataset}. The dataset is created in a very synthetic manner using  item categories and properties to generate user queries: the terms from each product's category are concatenated following their hierarchy order to create a topic string. Stopwords and duplicate terms are removed from the topic string that is then used as a query for the associated item.
When removing the duplicated words, the terms from a lower category are preserved, e.g., $Camera \to Photo \to Digital\ Camera\ Lenses$ is
converted to \textit{“photo digital camera lenses”}. Given an item $i$ purchased by a user $u$, the item is considered relevant for $u$, and the synthetic query is generated as explained above.
This process comes with some drawbacks: it generates a low number of unique queries that do not resemble real-world queries as rarely a user writes down all the categories of a product in a hierarchical order to search for it.

Table \ref{tab:pir_compare} shows that the proposed dataset is, in terms of corpus volume, very similar to the other datasets, and it has a comparable value also for the other statistics. Actually, it is the largest one in terms of number of queries provided.
In terms of relevance assessment, it is the only one that has been explicitly annotated by the users: by a single user for the best answer and by various community users for relevance, by either up-voting it if the answer was relevant according to them or down-voting it otherwise.

\section{Preliminary experiments with \ourdataset}
\label{sec:experiments}
In this section, we briefly describe the experimental setup and introduce the methods employed to showcase  \ourdataset 
on the cQA task defined in  Section \ref{subsec:taskdefinition}. Finally, we report and discuss the results of the preliminary experiments conducted.

\subsection{Experimental settings}

We adopt a two-stage ranking architecture aimed at trading-off  effectiveness and efficiency by applying two increasingly accurate and computationally expensive ranking models. The first stage is inexpensive and recall-oriented. It aims at selecting for each query a set of candidate documents that are eventually re-ranked by the second, precision-oriented ranker.
The first stage is based on elasticsearch, and uses  BM25 as a fast ranker. To increase the recall in the set of candidate documents retrieved by the first stage, we optimize BM25 parameters by performing a grid search driven by Recall at 100 on a subset of 5000 queries randomly sampled from the validation set.  The optimal values for \textit{b} and \textit{k1} found are 1 and 1.75, respectively.
For the second, precision-oriented stage,  we rely on a linear combination of the scores computed by BM25, a neural re-ranker based on a pre-trained language model, and, when used, a personalization model exploiting user history, represented by the tags used by the users.  In all the experiments the second stage re-ranks the top 100 results retrieved with BM25.

\subsubsection*{Neural models}
We use the following three neural models in the second stage: 
\begin{itemize}
    \item The first model is MiniLM\footnote{\url{https://huggingface.co/sentence-transformers/all-MiniLM-L6-v2}}. This model was trained and tuned using billions of training pairs, given the presence of StackExchange pairs in the training data, we use the model as it is, without any fine-tuning;

    \item The second one is DistilBERT\footnote{\url{https://huggingface.co/distilbert-base-uncased}}. In this case we fine-tune the model using all the training queries of \ourdataset. For each query, one positive document and two negative documents are randomly sampled, one from the list retrieved by BM25, and one in-batch random negative~\cite{random_negative}. We fine-tune DistilBERT for 10 epochs, with a batch size of 16 and a learning rate of $10^{-6}$ by using Triplet Margin Loss~\cite{sbert}, with a margin $\gamma = 0.5$.

    \item The third one is MonoT5~\cite{nogueira-etal-2020-document}. It is based on a T5~\cite{raffel-t5} re-ranker, which is fine-tuned on the MS MARCO passage dataset\footnote{\url{https://huggingface.co/castorini/monot5-small-msmarco-10k}}\textsuperscript{,}\footnote{\url{https://huggingface.co/castorini/monot5-base-msmarco-10k}}. We train two different versions of MonoT5: small and base. To further fine-tune MonoT5-small, we follow the same setting proposed in~\cite{nogueira-etal-2020-document}, i.e. batch size equal to $128$ and learning rate of $10^{-3}$. For each query, we sample one positive document and one negative document from the list retrieved by BM25. To fine-tune MonoT5-base, we reduce the batch size to $64$ due to hardware constraints. Also in this case the models are trained for a total of 10 epochs. Instead of fine-tuning the whole model, we rely on Adapter modules~\cite{pfeiffer-2023-adapters,pfeiffer-etal-2021-adapterfusion}, composed of two Feed-Forward layers: the first one is a down projection of the input vector into an intermediate dimension, which is followed by a non-linear activation function; the second one is an up projection to the dimension of the input vector. Following Karimi et al.~\cite{karimi2021compacter}, the intermediate dimension is set to 48.
    
\end{itemize}

For DistilBERT and the two T5 models, we rely on AdamW~\cite{adamw} as the optimizer. 
For reproducibility purposes, we set the random seed to 42 for training DistilBERT and 0 for training the T5 models.

\subsubsection*{Personalized TAG model for cQA} 

For a given answer $\textbf{a}$ produced in response to a query $\textbf{q}$ formulated by a user $\textbf{u}$, a personalization score is computed as explained below. As previously explained this score is linearly combined with the BM25 score and with the score produced by the neural re-ranker. 
Given a question $\textbf{q}$, asked by user $\textbf{u}$ at time $\textbf{t}$, let $T_{u,t}$ be the set of tags assigned by $\textbf{u}$ to all her/his questions posted before $\textbf{t}$ (including $\textbf{q}$). 
$T_{u,t}$ thus represents the interests of u as expressed in her/his previous interactions. 
The authors of the answers to query \textbf{q} do not have the possibility of tagging explicitly their answers, so for each answer, we consider  the tags associated with the answered questions. 
Specifically, given an answer $\textbf{a}$ from a user $\textbf{u'}$, we represent $\textbf{a}$ with the set $T_{u',t}$, i.e. the set of all the tags associated to the questions to which $u'$ answered before $\textbf{t}$ (excluding $\textbf{q}$). It is worth noting that in computing $T_{u',t}$  we do not consider the tags associated with the current question $\textbf{q}$ to avoid data leakage.
The TAG model assigns to each answer $\textbf{a}$, which has been retrieved for question $\textbf{q}$ in the first stage, the following score:
\begin{displaymath}
    s_a = \frac{|T_{u',t} \cap T_{u,t}|}{|T_{u,t}| + 1},
\end{displaymath}
where we add $1$ in the denominator as a smoothing factor, needed for cases where set $T_{u,t}$ is empty.
The rationale behind the proposed formula is that an answer is assigned a higher score if the question author shares similar interests (represented by means of tags s/he assigned to her/his asked questions) to the answerer.

\subsubsection*{Score combination}
The final ranking is obtained  by computing  the weighted sum of the normalized scores from the above single models, i.e., by using  the weights $\lambda_{BM25}, \lambda_{Neural},$ and $\lambda_{TAG}$, for the BM25, Neural (MiniLM, DistilBERT or T5), and TAG models, respectively,  with $\sum_i \lambda_i = 1$.
The  $\lambda$ values are optimized on the validation set by performing a grid search in the interval $[0, 1]$ with step $0.1$.

\subsubsection*{Evaluation Metrics}
We use P@1, NDCG@3,  NDCG@10, Recall@100,  and MAP@100 as our evaluation metrics.  
The cutoffs considered are  low as it is important to find the relevant results at the top of the ranked lists.
All the metrics are computed by using the \textit{ranx} library~\cite{bassani2022ranx, ranx_fusion}. 
We make our runs publicly available, so different metrics can be computed if necessary. 

\subsection{Experimental Results}
\label{subsec:exResults}

\begin{table}
    \centering
    \caption{Results for the cQA task on  Base \ourdataset.% The symbol * indicates a statistically significant improvement over the respective non-personalized method.
    }
    \resizebox{\linewidth}{!}{
    \begin{tabular}{l c c c c c c}
        \toprule
        Model & P@1\ssp & NDCG@3\ssp & NDCG@10\ssp & R@100 & MAP@100\ssp & $\lambda$\\
        \midrule
        BM25 &  0.330\ssp & 0.325\ssp & 0.359\ssp & 0.615 & 0.320\ssp & -\\
        BM25 + TAG & 0.355* & 0.349* & 0.383* & 0.615 & 0.342 & (.7;.3)\\
        BM25 + DistilBERT & 0.404\ssp & 0.400\ssp & 0.435\ssp & 0.615 & 0.389\ssp & (.3;.7)\\
        BM25 + DistilBERT + TAG & 0.422* & 0.415* & 0.448* & 0.615 & 0.402* & (.3;.5;.2)\\
        BM25 + MiniLM & 0.473\ssp & 0.459\ssp & 0.486\ssp & 0.615 & 0.443\ssp & (.1;.9)\\
        BM25 + MiniLM + TAG & 0.493* & 0.475* & 0.500* & 0.615 & 0.457* & (.1;.8;.1)\\
        % BM25 + T5-small & 0.448\ssp & 0.442\ssp & 0.471\ssp & 0.615 & 0.426\ssp & (.1;.9)\\
        BM25 + T5-small + TAG & 0.448\ssp & 0.442\ssp & 0.471\ssp & 0.615 & 0.426\ssp & (.1;.9;.0)\\
        BM25 + T5-base + TAG & \textbf{0.497}\ssp & \textbf{0.491}\ssp & \textbf{0.514}\ssp & 0.615 & \textbf{0.470}\ssp & (.1;.9;.0)\\
        % BM25 + T5-base & 0.497\ssp & 0.491\ssp & 0.514\ssp & 0.615 & 0.470\ssp & (.1;.9;.0)\\
        
    \end{tabular}
    }
    \label{tab:answer_base_result}
    
    \caption{Results for the cQA task on  Pers \ourdataset. %The symbol * indicates a statistically significant improvement over the respective non-personalized method.
    }
    \resizebox{\linewidth}{!}{
    \begin{tabular}{l c c c c c c}
        \toprule
        Model & P@1\ssp & NDCG@3\ssp & NDCG@10\ssp & R@100 & MAP@100\ssp & $\lambda$\\
        \midrule
        BM25 & 0.279\ssp & 0.353\ssp & 0.394\ssp & 0.707 & 0.362\ssp & -\\
        BM25 + TAG & 0.306* & 0.383* & 0.425* & 0.707 & 0.392* & (.7;.3)\\
        BM25 + DistilBERT & 0.351\ssp & 0.437\ssp & 0.478\ssp & 0.707 & 0.441\ssp & (.3;.7)\\
        BM25 + DistilBERT + TAG & 0.375* & 0.460* & 0.500* & 0.707 & 0.463* & (.3;.5;.2)\\
        BM25 + MiniLM & 0.403\ssp & 0.491\ssp & 0.525\ssp & 0.707 & 0.490\ssp & (.1;.9)\\
        BM25 + MiniLM + TAG & 0.426* & 0.512* & 0.543* & 0.707 & 0.509* & (.1;.8;.1)\\
        BM25 + T5-small & 0.376\ssp & 0.469\ssp & 0.506\ssp & 0.707 & 0.468\ssp & (.1;.9)\\
        BM25 + T5-small + TAG & 0.400* & 0.491* & 0.525* & 0.707 & 0.489* & (.1;.8;.1)\\
        BM25 + T5-base & 0.417\ssp & 0.517\ssp & 0.548\ssp & 0.707 & 0.510\ssp & (.1;.9)\\
        BM25 + T5-base + TAG & \textbf{0.440}* & \textbf{0.535}* & \textbf{0.563}* & 0.707 & \textbf{0.528}* & (.1;.8;.1)\\  
        
    \end{tabular}
    }
    \label{tab:answer_pers_result}
\end{table}

The results of the experiments conducted are reported in Table \ref{tab:answer_base_result} and  \ref{tab:answer_pers_result} for the \textit{base}  and \textit{pers} versions of the dataset, respectively. 
In all  tables, the symbol * indicates a statistically significant improvement over the respective non-personalized method not using any contribution from the TAG model. Statistical significance is assessed with a Bonferroni-corrected two-sided paired student's t-test with 99\% confidence. Moreover, in the column labeled $\lambda$ we report the optimized weights used for combining the scores computed by BM25, DistilBERT/MiniLM, and TAG models.
From the two tables,  we notice first that neural re-rankers are effective and that the methods using MiniLM outperform those based on DistilBERT and MonoT5-small. This  was somehow expected due to the huge  training  set used for training MiniLM~\cite{huggingface_all}. 
After fine-tuning, MonoT5-base, with more parameters, is able to achieve better results than MiniLM. 
Moreover, MonoT5-small and MonoT5-base, fine-tuned for just 10 epochs on the proposed dataset, improve by 33\% and 47\% in MAP@100, respectively, the BM25 performance.
However, the most notable result is that TAG improves, by a statistically significant margin, any cQA method it is combined with and for all the metrics considered on the \textit{pers} version of the dataset (Table \ref{tab:answer_pers_result}), where non-personalizable queries are removed, thus showing the advantages of personalization. 
The improvement due to the addition of this simple personalized model reaches up to 8\% in terms of MAP@100 compared to their non-personalized baseline.

\begin{table}
    \centering
    \caption{Results for the cQA task on  single-community data extracted from Base \ourdataset.}
    \resizebox{\linewidth}{!}{
    \begin{tabular}{l l c c c c c c}
        \toprule
        Community & Model (BM25 +)    & P@1   &    NDCG@3   &    NDCG@10  &   R@100\ssp  &  MAP@100 & $\lambda$ \\
        \midrule
        \multirow{2}{*}{Academia}
        & MiniLM & 0.438\ssp & 0.382\ssp & 0.395\ssp & 0.489\ssp & 0.344\ssp & (.1,.9)\\
        & MiniLM + TAG & \textbf{0.453}* & \textbf{0.392}* & \textbf{0.403}* & 0.489\ssp & \textbf{0.352}* & (.1,.8,.1)\\
        \midrule
        \multirow{2}{*}{Apple}
        & MiniLM & 0.327\ssp & 0.351\ssp & 0.381\ssp & 0.514\ssp & 0.349\ssp & (.1,.9)\\
        & MiniLM + TAG & \textbf{0.335}* & \textbf{0.361}* & \textbf{0.389}* & 0.514\ssp & \textbf{0.357}* & (.1,.8,.1)\\
        \midrule
        \multirow{2}{*}{Bicycles}
        & MiniLM & 0.405\ssp & 0.380\ssp & 0.421\ssp & 0.600\ssp & 0.365\ssp & (.1,.9)\\
        & MiniLM + TAG & \textbf{0.436}* & \textbf{0.405}* & \textbf{0.441}* & 0.600\ssp & \textbf{0.386}* & (.1,.8,.1)\\
        \midrule
        \multirow{2}{*}{Christianity}
        & MiniLM & 0.534\ssp & 0.505\ssp & 0.555\ssp & 0.783\ssp & 0.497\ssp & (.2,.8)\\
        & MiniLM + TAG & \textbf{0.549}* & \textbf{0.521}* & \textbf{0.564}* & 0.783\ssp & \textbf{0.507}* & (.1,.8,.1)\\
        \midrule
        \multirow{2}{*}{Cooking}
        & MiniLM        & 0.600\ssp  &    0.567\ssp   &   0.600\ssp    &   0.719\ssp     & 0.553\ssp    & (.1,.9)\\
        & MiniLM + TAG  & \textbf{0.619}*  &    \textbf{0.583}*   &   \textbf{0.614}*    &   0.719\ssp     & \textbf{0.568}*    &  (.1,.8,.1)\\
        \midrule
        \multirow{2}{*}{DIY}
        & MiniLM        & 0.323\ssp  &     0.313\ssp  &     0.346\ssp  &     0.501\ssp   &    0.302\ssp    & (.1,.9) \\
        & MiniLM + TAG  & \textbf{0.335}*  &     \textbf{0.324}*  &     \textbf{0.356}*  &     0.501\ssp   &    \textbf{0.312}* & (.1,.8,.1)\\
        \midrule
        \multirow{2}{*}{Hermeneutics}
        & MiniLM        & 0.589\ssp  &     0.538\ssp  &     0.593\ssp  &     0.828\ssp   &    0.526\ssp   &  (.2,.8)\\
        & MiniLM + TAG  & \textbf{0.632}*  &   \textbf{0.570}*  &  \textbf{0.617}*  & 0.828\ssp   & \textbf{0.552}*    & (.1,.8,.1)\\
        \midrule
        \multirow{2}{*}{Law}
        & MiniLM        & 0.663\ssp  &    0.647\ssp  &    0.678\ssp   &    0.803\ssp     & 0.639\ssp    & (.2,.8)\\
        & MiniLM + TAG  & \textbf{0.677}*  &    \textbf{0.657}*  &    \textbf{0.687}*   &    0.803\ssp     & \textbf{0.649}*  & (.1,.8,.1)\\
        \midrule
        \multirow{2}{*}{Money}
        & MiniLM        & 0.545\ssp  &     0.535\ssp  &     0.563\ssp  &     0.706\ssp   &    0.515\ssp   &  (.2,.8)\\
        & MiniLM + TAG  & \textbf{0.559}*  &   \textbf{0.542}*  &  \textbf{0.571}*  & 0.706\ssp   & \textbf{0.523}*    & (.1,.8,.1)\\
        \midrule
        \multirow{2}{*}{Music}
        & MiniLM        & 0.508\ssp  &     0.447\ssp  &     0.476\ssp  &     0.602\ssp   &    0.418\ssp   &  (.2,.8)\\
        & MiniLM + TAG  & \textbf{0.522}*  &   \textbf{0.460}*  &  \textbf{0.486}*  & 0.602\ssp   & \textbf{0.427}*    & (.1,.8,.1)\\
        \midrule
        \multirow{2}{*}{Rpg}
        & MiniLM        & 0.657\ssp  &     0.646\ssp &      0.685\ssp &      0.849\ssp    &   0.640\ssp   &  (.2,.8) \\
        & MiniLM + TAG  & \textbf{0.677}*  & \textbf{0.660}* & \textbf{0.695}* & 0.849\ssp    &   \textbf{0.651}*     & (.1,.8,.1)\\
        \midrule
        \multirow{2}{*}{Scifi}
        & MiniLM        & 0.532\ssp  &     0.563\ssp &      0.596\ssp &      0.745\ssp    &   0.559\ssp   &  (.2,.8) \\
        & MiniLM + TAG  & \textbf{0.549}*  & \textbf{0.574}* & \textbf{0.606}* & 0.745\ssp    &   \textbf{0.569}*     & (.1,.8,.1)\\
        \midrule
        \midrule
        $\lambda_{TAG} = 0$ & \multicolumn{7}{l}{\parbox{1.3\linewidth}{english, health, history, travel, workplace, writers, woodworking, vegetarianism, skeptics, politics, philosophy, parenting, outdoors, musicfans, literature, linguistics, judaism, interpersonal, hsm, genealogy, freelancing, fitness, expatriates, buddhism, anime}}\\
        \midrule
        \midrule
        \parbox{.22\linewidth}{No Statistical Improvement} & \multicolumn{7}{l}{\parbox{1.3\linewidth}{boardgames, gardening, gaming, hinduism, islam, lifehacks, martialarts, movies, opensource, pets, sound, sports, sustainability}}\\
        
 %       \midrule
 %        \midrule
  %       \multirow{2}{*}{Average}
  %       & MiniLM       & 0.519\ssp  &    0.506\ssp  &    0.536\ssp    &   0.677\ssp   &   0.492\ssp   & - \\ 
  %       & MiniLM + TAG & \textbf{0.530}*  &    \textbf{0.515}*  &    \textbf{0.544}*    &   0.677\ssp   &   \textbf{0.500}*   & -\\
        \bottomrule
    \end{tabular}
    }
    \label{tab:comparision_domain}
\end{table}

% no improvement for english, health, history, travel, workplace, writers, woodworking, vegetarianism, skeptics, politics, philosophy, parenting, outdoors, musicfans, literature, linguistics, judaism, interpersonal, hsm, genealogy, freelancing, fitness, expatriates, buddhism, anime -> 25 communities

% No statistical: Boardgames, Gardening, gaming (P@1), hinduism (decrease in performance), islam (P@1), Lifehacks, martialarts, Movies (P@1), opensource, Pets (P@1), sound, sports, Sustainability (P@1), 

Finally, in order to validate our hypothesis that personalization is more useful on a multi-domain, heterogeneous collection than on a single-domain, homogeneous  one, we perform a series of experiments considering single-domain data extracted from \ourdataset. Specifically, we consider 50 partitions of \ourdataset (only base version) built by isolating the data from the 50 communities. We apply to each one of these subsets the non-personalized and personalized combinations of models using the best performing bi-encoder model MiniLM, and measure the performance according to the same metrics used for the previous cQA tests. We did not employ the T5 models due to their long computation time. For a fair comparison, we performed for each community the optimization of the $\lambda$ weights on single-domain validation data. 
Differently from the multi-domain results shown in Table \ref{tab:answer_base_result}, we notice that the contribution of the TAG model is lower, and in some cases missing. Specifically, for 25 out of 50 communities, personalization does not lead to any improvement, i.e., $\lambda_{TAG} = 0$. On the other  13 communities, we do not observe statistically significant improvements for P@1 over the non-personalized methods. 
Since statistical significance is affected also by the size of the sample, we computed also the  performance metrics averaged on all the runs with single-domain data.  
As expected, the absolute metrics are slightly higher for single-domain tests due to the higher recall in the first-stage retrieval. 
In fact, by considering single-community  data at a time, we drastically reduce the size of the collection indexed, allowing the first-stage ranker to perform better.
However, in terms of the absolute performance boost due to the TAG model,  we achieve a 2\% improvement on P@1  when using all communities together, while the boost decreases to 1.1\% when considering the communities separately. 
The results of these experiments are reported integrally in the \ourdataset Zenodo and Github page. Here, in Table \ref{tab:comparision_domain}, we report the results for the 12 communities for which personalization achieves statistically significant improvements. 

\section{Utility and predicted impact}
\label{sec:utility} 
The \ourdataset resource we make available to the research community  is a step ahead toward a fair and robust evaluation of personalization approaches in Information Retrieval.  The features provided with the dataset include explicit signals to create relevance judgments and a large amount of historical user-level information  allowing to design  and test classical and novel personalization methods.
We expect the \ourdataset dataset to be useful for many researchers and practitioners working in personalized IR and in the application of machine/deep learning techniques for personalization. In recent years, the IR community spent important effort in studying personalization. However, a comprehensive dataset for evaluating and comparing different approaches is still missing. Researchers  mainly  rely on synthetic datasets or use non-public data, which makes the comparison between different methods less reliable or, worse, not possible at all. The \ourdataset  dataset advances this  research area by filling this gap with a large-scale dataset covering the activity of StackExchange users in a period of 14 years. For this reason, we expect that the dataset will impact the  research community working on personalized IR as it provides a single common ground of evaluation built on   questions \& answers from real users socially interacting via a community-oriented  web platform.

\section{Conclusion and Future Works}
\label{sec:conclusion}

This paper discussed the characteristics of \ourdataset (StackExchange - Personalized Question Answering), a large real-world dataset including about 1 million questions and 2 million associated answers contributed by  the users of StackExchange communities.  The data comes with a rich set of  user-level features modeling the  interactions  among the members of the online communities, e.g., the  positive or negative votes received by questions and answers, the tags associated with questions, the comments that other users might have written under a question or an answer, the  users' autobiographies, reputation score, and the number of views  received by their profile.

We detailed all the information available in the dataset and discussed how it can be  exploited  for training and evaluating classical and personalized models addressing \textit{cQA task}. As exemplifying methodologies, we focused on IR approaches for these tasks based on a two-stage architecture where the second re-ranking stage exploits a combination of the scores computed by BM25, DistilBERT/MiniLM/T5, and TAG models.
The results of the preliminary experiments conducted show that personalization works effectively on this dataset, improving by a statistically significant margin, in most of the cases, state-of-the-art methods based on pre-trained large language models.  

The analysis conducted and the peculiarities of  the \ourdataset resource suggest several lines of future investigation. 
For example, in this work we employed a relatively simple user model for personalization, we leave the development of more complex personalized models for future works that could exploit user features of \ourdataset that were not used in the proposed models.

\bibliographystyle{ACM-Reference-Format}
\bibliography{biblio}

\end{document}